\documentclass[twocolumn,superscriptaddress,amssymb,amsmath,nobibnotes,aps,prl]{revtex4}

\usepackage{color}
\usepackage{graphicx}

\newcommand{\ee}{{\mathrm e}}
\newcommand{\ii}{{\mathrm i}}

\newcommand{\zz}{{\mathbb Z}}
\newcommand{\Tr}{\operatorname{Tr}}
\newcommand{\ndel}[1][]{{\partial^{#1} \!}}
\newcommand{\ndif}[1][]{{\mathrm{d}^{#1} \!}}
\newcommand{\dif}[2][]{{\ndif[#1] #2 \,}}
\newcommand{\cpval}{\mathcal P}

\begin{document}

\title{The entanglement spectrum of chiral fermions on the torus}

\author{Pascal Fries}
\email{pascal.fries@physik.uni-wuerzburg.de}
\affiliation{Fakult\"at f\"ur Physik und Astronomie, Julius-Maximilians Universit\"at
  W\"urzburg, Am Hubland, 97074 W\"urzburg, Germany}
\author{Ignacio A. Reyes}
\email{ignacio.reyes@aei.mpg.de}
\affiliation{Max-Planck-Institut f\"ur Gravitationsphysik (Albert Einstein Institute), Am M\"uhlenberg 1, 14476
  Potsdam, Germany}

\date{\today}

\begin{abstract}

  We determine the modular Hamiltonian of chiral fermions on the torus, for an arbitrary
  set of disjoint intervals at generic temperature. We find that, in addition to a local
  Unruh-like term, each point is non-locally coupled to an infinite but discrete set of
  other points, even for a single interval. These accumulate near the boundaries of the
  intervals, where the coupling becomes increasingly redshifted. Remarkably, in the
  presence of a zero mode, this set of points ``condenses'' within the interval at low
  temperatures, yielding continuous non-locality.
\end{abstract}

\maketitle

\section{Introduction}
Amongst the predictions stemming from the interplay between Quantum Field Theory (QFT) and
the causal structure of spacetime, one of the most robust is the celebrated Unruh effect:
An accelerated observer in the vacuum measures a thermal bath, with a temperature
proportional to its proper
acceleration~\cite{Fulling:1972md,Davies:1974th,Unruh:1976db}. Intimately connected with
the thermodynamics of black holes via Hawking radiation, this lies at the heart of our
current understanding of the quantum nature of gravity~\cite{Hawking:1978jn}. Therefore,
it is natural to explore its generalisations and investigate it further.

In recent years, these phenomena have been extended into the framework of quantum
information theory. There, this temperature is understood as arising from the entanglement
structure of the vacuum. Starting from a state $\rho$ and some entangling subregion $V$,
one defines the reduced density matrix $\rho_V$ by tracing out the complement of
$V$. Then, just as the entanglement entropy $S_V=-\Tr[\rho_V \log \rho_V]$ generalises the
thermal entropy, the usual Hamiltonian is an instance of the more general concept of a
\emph{modular} (or entanglement) Hamiltonian $\mathcal{K}_V$ defined via
\begin{align}\label{}
\rho_V:=\frac{e^{-\mathcal{K}_V}}{\text{tr}\, e^{-\mathcal{K}_V} }
\end{align}

Originally introduced within algebraic QFT~\cite{Haag:1992hx}, the modular Hamiltonian has
aroused much interest across a wide community due its close connection to quantum
information measures. In the context of many body quantum systems, the spectrum of this
operator is known as the ``entanglement spectrum'' and has been proposed as a fingerprint
of topological order~\cite{Li:2008kjh,Chandran:2011sdf,Dalmonte:2017bzm} and investigated
in lattice
models~\cite{Peschel:2009zgv,Eisler:2018zsf,Zhu:2018wsx,Parisen:2018gdj,Luitz:2014zgv}, as
well as tensor networks~\cite{Cirac:2011iuz,Hsieh:2014jba,Vanderstraeten:2017unh}. In QFT,
it is fundamental for the study of relative entropy~\cite{Sarosi:2017rsq,Casini:2017roe}
and its many applications to energy and information
inequalities~\cite{Casini:2008cr,Blanco:2013lea,Faulkner:2016mzt}. In the context of the
AdS/CFT correspondence, it is instrumental in the program of reconstructing a
gravitational bulk from the holographic
data~\cite{Casini:2011kv,Blanco:2013joa,Jafferis:2014lza,Jafferis:2015del,Lashkari:2013koa,Koeller:2017njr,Chen:2018rgz,Belin:2018juv,Abt:2018zif,Jefferson:2018ksk}.

However, the modular Hamiltonian is known in only a handful of cases. The result is
universal and local for the vacuum of any QFT reduced to Rindler
space~\cite{Unruh:1976db,Bisognano:1976za} and hence any CFT vacuum on the plane reduced
to a ball~\cite{Casini:2011kv}. For any CFT$_2$, the same applies for a single interval,
for the vacuum on the cylinder or a thermal state on the real
line~\cite{Hartman:2015apr,Cardy:2016fqc}. More generically, modular flows can be
non-local, as is the case for multiple intervals in the vacuum of chiral fermions on the
plane or the cylinder~\cite{Casini:2009vk,Klich:2015ina} and scalars on the
plane~\cite{Arias:2018tmw}.  The exact nature of the transit from locality to non-locality
however is not fully understood, and remains an active topic of research.

In this paper we report progress regarding this problem, by providing a new entry to this
list. We show that the chiral fermion on the torus (finite temperature on the circle) is a
solvable model that undergoes such a transition between locality and non-locality. We
compute the modular Hamiltonian by restating the problem as a singular integral
equation, which in turn we solve via residue analysis.

  Let us quickly quote our main result. For generic temperature, the modular Hamiltonian
  takes the form
  \[
    \mathcal{K}_{\text{loc}} + \mathcal{K}_{\text{bi-loc}}.
  \]
  The local flow is of the standard Rindler form \eqref{eq:unruh}, with entanglement
  temperature given in \eqref{eq:beta}. The novel result is the second term, given in
  \eqref{eq:Kpmbiloc} and depicted in Fig.~\ref{fig:mu-roots}, involving bi-local
  couplings between a discrete but infinite set of other points within the subregion. In
  the low temperature limit, the sector with a zero mode experiences a ``condensation'' of
  these points, resulting in a completely non-local flow.

\section{The resolvent}
We start by introducing the resolvent method,
following~\cite{Casini:2009vk,Klich:2017qmt,Arias:2018tmw}. For any spatial region $V$,
the reduced density matrix $\rho_V$ is defined as to reproduce expectation values of local
observables supported on $V$. Now, for free fermions, Wick's theorem implies that it is
sufficient that $\rho_V$ reproduces the equal-time Green's function
\begin{equation*}
  \Tr[\rho_V \psi^{}(x)\psi^\dagger(y)] =
  \langle \psi^{}(x)\psi^\dagger(y) \rangle =: G(x,y)
\end{equation*}
for $x,y\in V$. This requirement fixes the modular Hamiltonian to be a quadratic operator given by~\cite{Peschel:2003xzh}
  \begin{equation}
    \label{rho}
    \mathcal K_V = \int_V \!\dif x \int_V \!\dif y K_V(x,y) \psi^\dag(x) \psi(y)
  \end{equation}
  with kernel $K_V = -\log[G|_V^{-1} - 1]$. This is specific for
  the free fermion. $G_V$ refers to the propagator as the \emph{kernel of an
    operator} acting on functions with support on $V$.

  As shown in~\cite{Casini:2009vk} the modular Hamiltonian can be rewritten as
  \begin{equation}
    \label{eq:mod-ham-int}
    K_V = -\int_{1/2}^\infty \dif \xi [R_V(\xi)+R_V(-\xi)]
  \end{equation}
  in terms of the \emph{resolvent} of the propagator,
  \begin{equation}
    \label{eq:def-resolvent}
    R_V(\xi) := (G|_V + \xi - 1/2)^{-1}.
  \end{equation}

  A derivation of ~\eqref{eq:mod-ham-int} is provided in the
  supplemental material. In essence, it is the operator version of
  \[
    \log X = \frac{1}{2\pi i} \oint_\gamma \dif z \frac{\log z}{z-X} 
  \]
  for a suitable choice of contour $\gamma$.
  
  In \eqref{eq:def-resolvent}, the inverse of an operator is understood
  in the sense of a kernel,
  \[
    \int_V \!\dif z R_V(\xi;x,z) \Big[G(z,y)+(\xi-1/2)\delta(z,y)\Big] = \delta(x-y).
  \]
  Thus, provided $G$ of the global state and the entangling region $V$,
  this equation completely determines the resolvent $R_V$ and hence the modular
  Hamiltonian via \eqref{eq:mod-ham-int}.

To obtain the resolvent, let us first do the redefinition
\begin{equation}
  \label{eq:resolvent-ansatz}
  R_V(\xi;x,y) = \frac{\delta(x-y)}{\xi-1/2} - \frac {F_V(\xi;x,y)}{(\xi-1/2)^2}.
\end{equation}

  The convenience of this is that the first term of~\eqref{eq:resolvent-ansatz}
  will cancel the RHS of the previous equation, 
 translating~\eqref{eq:def-resolvent} into a singular integral equation
\begin{align}
  &0 = G(x,y) - F_V(\xi;x,y) \nonumber \\
  &\hspace{2.3cm} - \frac 1{\xi-1/2} \int_V \dif z G(x,z)F_V(\xi;z,y). \label{eq:singular-int-eq}
\end{align}

All previous considerations hold for free fermions on a generic Riemann surface. The
simplest case is the plane where the solution of~\eqref{eq:singular-int-eq} is a standard
result~\cite{Muskhelishvili:1958zgf}, which was used by~\cite{Casini:2009vk} to derive the
corresponding modular Hamiltonian. They found that for multiple intervals, it consists of
a local and a bi-local term. The former can be written as
\begin{equation}
  \label{eq:unruh}
  \mathcal K = \int_V \dif x \beta(x) T(x)
\end{equation}
in terms of the stress tensor 
  $T = \frac{\ii}{2} [\psi^\dag \partial_x \psi-\psi \partial_x \psi^\dag]$, where
$\beta(x)$ is known as the \emph{entanglement temperature}. On the other hand, the
bi-local term couples the field between different intervals.

Let us now proceed to the case of a chiral fermion on the torus. As is customary, we take
the periods to be $1,\tau$ with $\Im(\tau) > 0$, such that the nome $q:= \ee^{\ii\pi\tau}$
is inside the unit disk. We restrict to purely imaginary modulus $\tau=i\beta$, where $\beta$ is
the inverse temperature -- the general case can be recovered by analytic
continuation. For simplicity, we move to radial coordinates $w = \ee^{\ii\pi z}$.

Since we are dealing with fermions, the correlator $G(u,v)$ with $u = \ee^{\ii\pi x}$ and
$v = \ee^{\ii\pi y}$ is either periodic (Ramond; R) or anti-periodic (Neveu-Schwarz; NS)
with respect to either of the two periods of the torus. We shall restrict to
  the ``thermal'' case, with NS periodicity with respect to $\tau$. Combining this with the
requirement to reproduce the UV correlator $G^{\text{UV}}(x,y) = [2\pi\ii(x-y)]^{-1}$ on
small scales, this fully determines the standard Green's functions~\cite{DiFrancesco:1997nk}
\begin{equation}
  \label{eq:correlator}
  G^\nu(u,v) =
  \frac{\eta^3(q^2)}{\ii \vartheta_1(uv^{-1}\ee^\epsilon|q)}
  \frac{\vartheta_\nu(uv^{-1}|q)}{\vartheta_\nu(1|q)},
\end{equation}
where $\eta(q)$ and $\vartheta_\nu(z|q)$ are the Dedekind eta and Jacobi theta functions (see supplemental material).

Here, the superscript
\begin{equation*}
  \nu = 2,3 = (\text{R},\text{NS}), (\text{NS},\text{NS})
\end{equation*}
labels the different spin-structures, and we introduced a regulator $\epsilon$ in order to
treat the distribution $G^\nu$ as a function. The sign of $\epsilon$ depends on the
chirality---without loss of generality, we choose $\epsilon>0$.

  With the notation settled, we now go back to the integral
  equation~\eqref{eq:singular-int-eq}. In radial coordinates, it reads
\begin{align}
  &0 = G^\nu(u,v) - F^\nu_V(\xi;u,v) \nonumber \\
  &\hspace{1.3cm} - \frac 1{\xi-1/2} \frac 1{\ii\pi} \int_A
    \frac{\ndif w}w G^\nu(u,w)F^\nu_V(\xi;w,v) \label{eq:singular-int-eq-rad}
\end{align}
with $A := \ee^{\ii\pi V}$ being the entangling region. The key observation of this paper is that~\eqref{eq:singular-int-eq-rad}
  resembles the result of a contour integral, involving simple poles and branch cuts. Thus
  the strategy to solving~\eqref{eq:singular-int-eq-rad} is to recast it as a contour
  integral.  

\begin{figure}[h]
    \begin{center}
      \includegraphics[width=7.0cm]{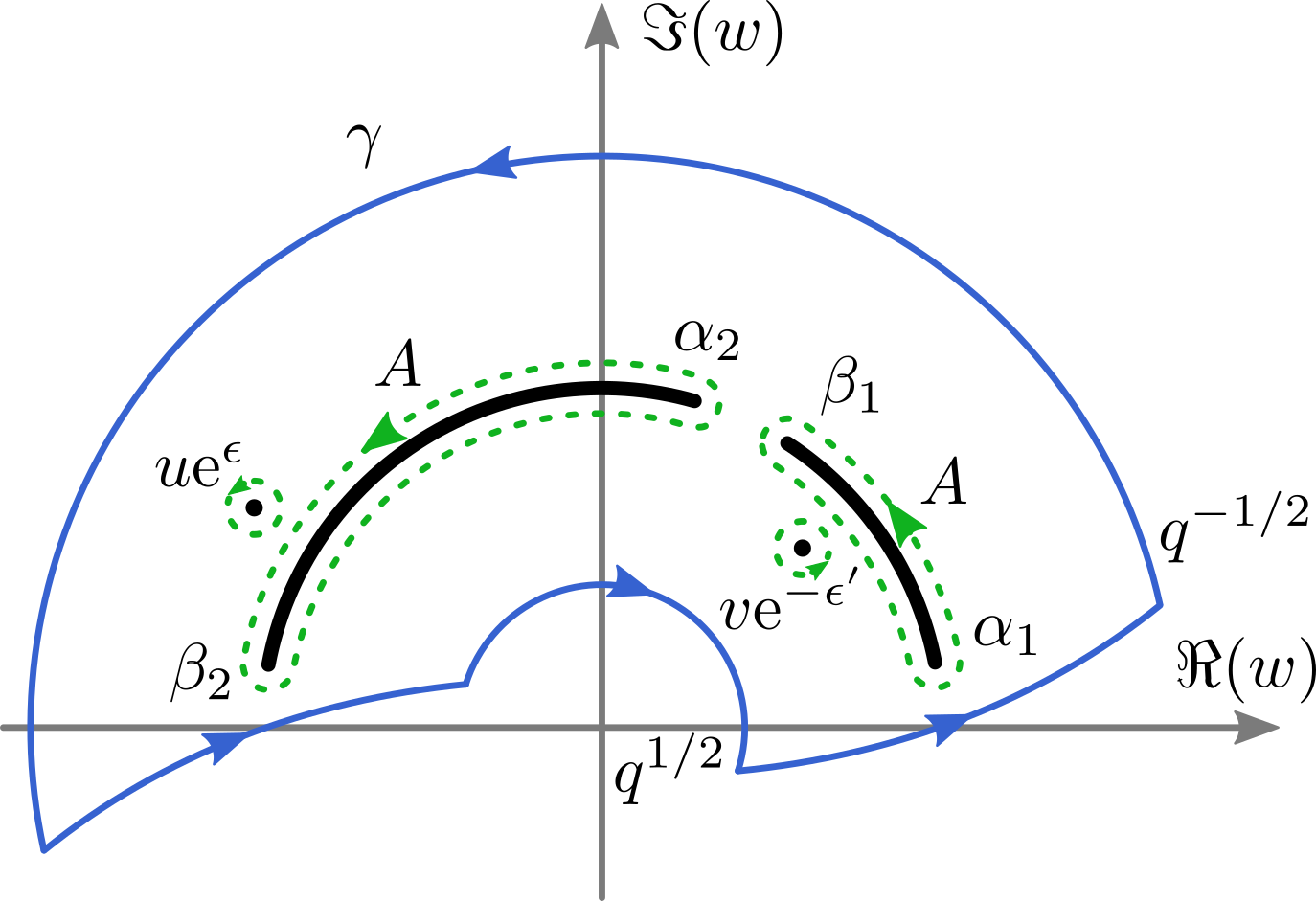} 
      \caption{ The complex plane analysis in the argument. The black solid line is the
        entangling region---here for simplicity two intervals. The blue line
        represents the contour of integration $\gamma$ in~\eqref{eq:gamma}, which leads to the
        residues evaluated along the green dotted curves.}
      \label{fig:gamma}
    \end{center}
  \end{figure}

\pagebreak
To this end, we start by listing a set of sufficient properties that $F^\nu_V$ must
possess in order to solve this equation:

  \textbf{(A) Periodicities.} First, it must have the same periodicities in
  the $w$ argument as $G^\nu$, such that $G^\nu F^\nu_V$ is well defined on the torus. The reason is that doubly periodic functions have vanishing residue along the boundary $\gamma$ of any
  fundamental region (see Fig.~\ref{fig:gamma}):
  \begin{equation}
    \label{eq:gamma}
    0 = \frac 1{\ii\pi} \oint_\gamma \frac{\ndif w}w G^\nu(u,w)F^\nu_V(\xi;w,v).
  \end{equation}
  Our aim is now to rewrite this in the form of~\eqref{eq:singular-int-eq-rad}. 

  \textbf{(B) Location of poles and branch cuts.}  The next property we demand is that
  $F^\nu_V$ have a simple pole $F^\nu_V(u,v) \sim 1/2(uv^{-1}-1)$ at $u \to v$, together
  with a branch cut along the entangling region $A$, which we specify below. Everywhere
  else it must be analytic. Note that, similarly to $G^\nu$, we need to introduce a
  regulator $\epsilon'>0$ for the pole of $F^\nu_V$.

  If these conditions are met, a simple residue analysis shows that~\eqref{eq:gamma} reduces to
  \begin{align}
    0 &= G^\nu(u,v\ee^{-\epsilon'}) - F^\nu_V(\xi;u\ee^\epsilon,v\ee^{-\epsilon'}) \nonumber \\
      &\hspace{.3cm} - \frac 1{\xi-1/2} \frac 1{\ii\pi} \int_{A^\circlearrowleft}\!\!
        \frac{\ndif w}w G^\nu(u\ee^\epsilon,w)F^\nu_V(\xi;w,v\ee^{-\epsilon'}),\label{eq:snug}
  \end{align}
  where we made the regulators explicit and $A^\circlearrowleft$ denotes a snug path
  around the cut on $A$ as depicted in Fig.~\ref{fig:gamma}.

  \textbf{(C) Residues.} This last integral decomposes into three contributions: one along
  $A$ just inside the unit circle, one along $A$ just outside the unit circle, and
  contributions from the boundary points $\alpha_n=e^{i\pi a_n},\beta_n=e^{i\pi b_n}$ of
  $A$ as can be seen from Fig.~\ref{fig:gamma}. Our final requirements on $F_V^\nu$ are
  that the residues at $\ndel A$ vanish, while $F^\nu_V$ has to have a multiplicative
  branch cut along $A$: at every point, the ratio of the
  function just above and below the cut is a fixed number
  \begin{equation}
    \label{eq:branch-cut}
    \frac{F^\nu_V(u\ee^{-\epsilon''},v)}{F^\nu_V(u\ee^{+\epsilon''},v)} =
    \frac{\xi+1/2}{\xi-1/2} =: \ee^{2\pi h}.
  \end{equation}

  The solution to \eqref{eq:branch-cut} in the plane is 
  familiar: $F(z)=z^m$ with $m\not\in \mathbb Z$ possesses such a
  cut. Below we find the analogue of this on the torus.

  If properties \textbf{(A),(B),(C)} are satisfied, it is easy to show that such an $F_V$
  indeed solves the problem: our contour equation \eqref{eq:snug} becomes exactly the
  original singular integral equation~\eqref{eq:singular-int-eq-rad}. The requirement that
  the residues on $\ndel A$ vanish is equivalent to demanding that the modular flow
  behaves like Rindler space in the vicinity of $\ndel A$. This is analogous to the
  derivation of the black hole temperature by the smoothness condition at the
  horizon.

In the supplemental material, we explicitly derive $F^\nu_V$ satisfying all of the above
assumptions. The general procedure is as follows:
\begin{enumerate}
\item Start with the standard solution for the requirement of a multiplicative branch
  cut~\eqref{eq:branch-cut} on the cylinder~\cite{Klich:2015ina}.
\item Average over all fundamental domains in the direction of $\tau$. This yields a
  quasiperiodic function.
\item Multiply with a slightly modified form of the Green's function~\eqref{eq:correlator}
  to turn the quasiperiodicity into a periodicity and introduce the correct pole.
\end{enumerate}

We are now in position to state one of the main results of this paper: the resolvent
for a finite union of disjoint intervals on the torus, $V=\cup_{n=1}^N (a_n,b_n)$. The
exact expression lives in the complex plane, but is vastly simplified along
$A$. Introducing the shorthand notation
\begin{equation}
  \label{eq:def-lambda}
  \lambda
  := \bigg[\prod_{n=1}^N \frac{\alpha_n}{\beta_n}\bigg]^{\ii h} \!\!\!\!= \ee^{\pi h L},
\end{equation}
where $L$ is the total length of $V$, our result is
\begin{align}
  F^\nu_V(\xi;u,v)
  &= \frac{\eta^3(q^2)}{\ii \vartheta_1(uv^{-1}\ee^{\epsilon'}|q)}
    \frac{\vartheta_\nu(\lambda uv^{-1}|q)}{\vartheta_\nu(\lambda|q)} \nonumber \\
  &\hspace{2.5cm} \times \ee^{-2\pi h} \bigg[\frac{\Omega_V(u)}{\Omega_V(v)}\bigg]^{\ii h}
    \label{eq:solution}
\end{align}
with $h$ defined in~\eqref{eq:branch-cut}, and
\begin{equation}
   \label{eq:def-omega}
   \Omega_V(w)
   := - \prod_{n=1}^N \frac{\vartheta_1(w\alpha_n^{-1}|q)}{\vartheta_1(w\beta_n^{-1}|q)}.
\end{equation}

Some comments are in order. The term in the second line of \eqref{eq:solution} is the complex power of a quotient, which introduces the required
branch cut along $A$. This function is quasi-periodic, acquiring a factor of
$\lambda^2$ when translated into the next fundamental domain. The first factor resembles
the propagator~\eqref{eq:correlator} and introduces the desired pole, as described
above. Additionally, the extra factor of $\lambda$ in the argument of $\vartheta_\nu$ is
there to precisely cancel the quasi-periodicity of the second term. This allows the
product $G^\nu F^\nu_V$ to be exactly doubly periodic, as required.

\section{Modular Hamiltonian}
Finally, now that we have found the resolvent $R^\nu_V$, we can go back to~\eqref{eq:mod-ham-int}
to obtain the modular Hamiltonian $K^\nu_V$. First, note that the leading divergence of
$F^\nu_V(u,v) \sim 1/2(uv^{-1}\ee^{\epsilon'}-1)$ at $u \to v$ can be rewritten as a
Cauchy principle value
\begin{equation}
  \label{eq:cpval}
  \frac 12 \frac 1{uv^{-1}\ee^{\epsilon'}-1} =
  \frac{\delta(x-y)}2 + \cpval \frac 12\frac 1{uv^{-1}-1}.
\end{equation}

For the sake of readability, we shall keep $\cpval$ implicit for the rest of this
paper. Equation~\eqref{eq:cpval} implies that the $\delta$-terms
from~\eqref{eq:resolvent-ansatz} drop out in~\eqref{eq:mod-ham-int}, yielding
\begin{equation}
  \label{eq:mod-ham-int-2}
  K^\nu_V =
  \int_{1/2}^\infty \frac{\dif \xi}{(\xi-1/2)^2} \Big[F^\nu_V(\xi) + F^\nu_V(-\xi)\Big].
\end{equation}

The main characteristic of \eqref{eq:mod-ham-int-2} is that the integrand is highly oscillatory and divergent around
  $\xi=1/2$. Indeed, notice that when $\xi\to 1/2$ the prefactor
  in~\eqref{eq:mod-ham-int-2} diverges quadratically while $F(\xi)$ vanishes linearly but
  oscillates wildly due to the last factor in~\eqref{eq:solution}. However, this behaviour
  is well understood in the theory of distributions, and in this sense the
  expression~\eqref{eq:mod-ham-int-2} is well defined and
  closely related to the Dirac delta.

  In the supplemental material, we evaluate~\eqref{eq:mod-ham-int-2} analytically. Here we will simply quote the result, but
  the main steps in the derivation are the following:
  \begin{enumerate}
  \item Change variables to isolates all the infinite poles along the
    negative axis, which then lie in successive fundamental domains.
  \item Regularize~\eqref{eq:mod-ham-int-2} by placing a contour that includes
    increasingly many poles, and express it by residues.
  \item Use the quasiperiodicites of $\vartheta_\nu$ to bring every pole to the
    fundamental region, expressing~\eqref{eq:mod-ham-int-2} as a highly oscillatory
    function with a divergent prefactor.
  \item Remove the regulator, leading to standard Dirichlet kernel representations of the
    periodic/antiperiodic Dirac delta.
\end{enumerate}

The final expression for the modular Hamiltonian depends on the spin sector. Let us focus
on the results for a single interval. Both sectors $\nu=2,3$ have a local and a bi-local
term. The local term is identical in both cases and takes the form
\begin{equation}
  \label{eq:k-loc-23}
  K_{\text{loc}}(x,y) = \beta(x) [\ii\partial_x+ f(x)]\delta (x-y),
\end{equation}
with the entanglement temperature
\begin{equation}
  \label{eq:beta}
  \beta(x) = \frac{2\pi\beta}{2\pi + \beta\partial_x\log\Omega_V(\ee^{\ii\pi x})},
\end{equation}
where $\Omega_V$ is as defined in~\eqref{eq:def-omega} and the function $f(x)$ is fixed by
requiring that $K_{\text{loc}}$ is hermitian. Note that the expression~\eqref{eq:k-loc-23}
is equivalent to the more familiar Rindler-like representation~\eqref{eq:unruh}.

The bi-local term represents the central result of this paper and shows a novel feature:
In both sectors, it involves a coupling between an infinite but discrete set of points,
and is given by
\begin{align}
  &K^\pm_{\text{bi-loc}}(x,y) =\frac{\ii\pi}{L\sinh \pi\mu(x,y)} \nonumber\\
  &\hspace{1.5cm}\times \!\!\sum_{k\in \zz\setminus\{0\}}\!\!\! (\pm 1)^k
    \delta(x-y+\beta\mu(x,y)-k), \label{eq:k-biloc}
\end{align}
where the sign $\pm$ corresponds to $\nu=\genfrac{}{}{0pt}{1}{2}{3}$. Here, we used the function 
\begin{equation}
  \label{eq:def-mu}
  \mu(x,y) =\frac{1}{2\pi L} \log \frac{\Omega_V(\ee^{\ii\pi x})}{\Omega_V(\ee^{\ii\pi y})},
\end{equation}
which will play an important role in the analysis below.

\begin{figure}[h]
\begin{center}
  \includegraphics[width=7.3cm]{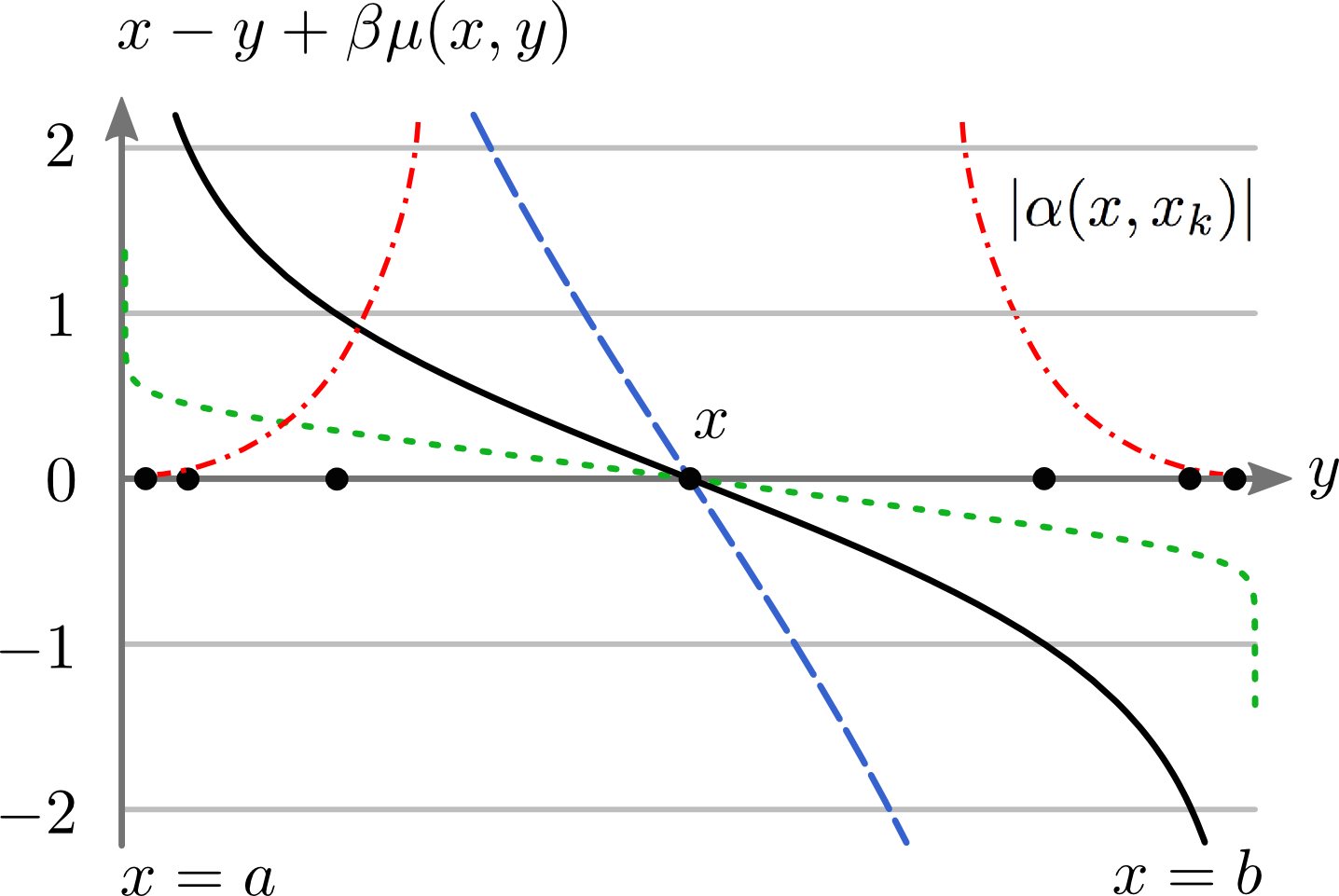}
  \caption{For finite $\beta$ (black solid) the point at the centre is bi-locally coupled
    to an infinite set $x_k(x)$ (black dots), solutions to \eqref{eq:mu-roots} for a
    single interval. For large $\beta$ (blue dashed), the solutions distribute densely,
    whereas for $\beta\to 0$ (green dotted) they all localise at the endpoints. The
    strength $\alpha(x,x_k)$ of the coupling (red, dot-dashed) decays towards the
    endpoints.}
  \label{fig:mu-roots}
\end{center}
\end{figure}

Note that $K^\pm_{\text{bi-loc}}$ couples pairs $(x,y)$ which are solutions of
\begin{equation}
  \label{eq:mu-roots}
  x-y+\beta\mu(x,y)-k=0, \quad k \in\zz\setminus\{0\}.
\end{equation}
Since $\mu(x,y)$ is monotonic in $y$ and diverges at the endpoints,
eq.~\eqref{eq:mu-roots} possesses a unique solution for every $k$, as shown in
Fig.~\ref{fig:mu-roots}. Solutions accumulate near the endpoints. In the next section, we
analyse the above expressions and discuss their physical meaning. A summary of the results
is presented in table~\ref{tab:summary}.

\section{Discussion}

In this paper we computed the modular Hamiltonian of chiral fermions in a
  thermal state on the circle, reduced to an arbitrary set of disjoint intervals.

  Our main result is that for arbitrary temperature, the modular Hamiltonian contains a
  local term, as well as an infinite number of bi-local contributions, even for a single
  interval. Let us now analyse the bi-local terms in more detail. Inserting the
  kernel~\eqref{eq:k-biloc} back into~\eqref{rho}, the bi-local modular Hamiltonian reads
  \begin{equation}
    \label{eq:Kpmbiloc}
    \mathcal{K}^\pm_{\text{bi-loc}}
    = \sum_{k\neq 0} (\pm 1)^k \int_V \!\dif x \alpha(x,x_k) \psi^\dag(x) \psi(x_k(x)).
  \end{equation}
  As depicted in Fig.~\ref{fig:mu-roots}, the $x_k(x)$ are an infinite set of points
  within the interval, solutions to equation~\eqref{eq:mu-roots}. The bi-local coupling
  $\alpha(x,x_k)$ has dimensions of energy and is given by
  \begin{align}
\alpha(x,y)=\frac{i\pi}{L\sinh \pi\mu(x,y)} \frac{1}{|1-\beta \partial_y\mu(x,y)|}\, .\nonumber
\end{align}
  
  Although determining the exact location of the $x_k$ is difficult, two properties are simple to
  extract:

  \textbullet~ The infinite set of $x_k$ accumulate near the endpoints of the
  interval. Indeed, since $\mu$ diverges there, there is an infinite number of solutions
  near the boundaries, located at
\begin{align}\label{x_k}
    x_k = a + e^{-2\pi L k/\beta}, \quad \text{as } k \to \infty
\end{align}

  and similarly near $b$.

  \textbullet~ Their contributions vanish as they approach the endpoints. Using~\eqref{eq:k-biloc}, the coupling in~\eqref{eq:Kpmbiloc} goes as
\begin{align}\label{alphaxxk}
    |\alpha(x,x_k)| \xrightarrow{k\to\infty} \frac{4\pi^2}\beta (x_k-a)^{1+1/2L}
\end{align}

 The energy scale of $\alpha(x,x_k)$ is set by the temperature
  $\beta^{-1}$, whereas the fall-off is determined by the length of the interval $L$. Interestingly, the strength of the
  non-local couplings appears to be ``redshifted'' due to their proximity to the local
  Rindler horizons located at the endpoints.

As a next step, let us see how to recover the known results at very
high~\cite{Cardy:2016fqc} and low~\cite{Klich:2015ina} temperatures. We start with the
high temperature limit $\beta\to 0$. One easily sees from~\eqref{eq:beta} that the local
term goes as the inverse temperature, $\beta(x)\sim\beta$, as expected. On the other hand,
as depicted in Fig.~\ref{fig:mu-roots}, the bi-local contributions~\eqref{eq:k-biloc} all
approach the endpoints, where they vanish exponentially.

Moving now to the low temperature limit $\beta\to\infty$, the entanglement
temperature~\eqref{eq:beta} approaches the well known result for the
cylinder~\cite{Klich:2015ina}
\begin{equation}
  \label{eq:beta-large-beta}
  \lim_{\beta\to\infty} \beta(x) =
  \frac{2\pi}{\partial_x \log \frac{\sin(x-a)}{\sin(b-x)}}.
\end{equation}

The bi-local contributions however behave remarkably. As can be understood from
Fig.~\ref{fig:mu-roots}, as we lower the temperature, the curve gets increasingly
steep. Thus, the solutions to~\eqref{eq:mu-roots} form a partition of the interval which
becomes denser and denser in the limit $\beta\to\infty$. Now, recall that the modular
Hamiltonian must always be thought of as a distribution, i.e. as integrated against
regular test functions. In this limiting procedure, the solutions to~\eqref{eq:mu-roots}
``condense'' in the interval, and it can be shown that the sequence of Dirac deltas
in~\eqref{eq:k-biloc} reproduce precisely the definition of a Riemann integral. Indeed,
one can show that in this sense~\eqref{eq:k-biloc} becomes completely non-local
\begin{equation}
  \label{eq:k-nonloc-large-beta}
  \lim_{\beta\to \infty} K^+_{\text{bi-loc}}(x,y) = \frac{\ii\pi}{L\sinh\pi\mu(x,y)},
\end{equation}
in agreement with~\cite{Klich:2015ina}, whereas
$\lim_{\beta\to\infty} K^-_{\text{bi-loc}} = 0$ due to the oscillating $(-1)^k$.

The previous analysis provides a new insight into the structure of fermionic entanglement: At any
finite temperature, non-locality couples a given point only to an infinite but discrete
set of other points. The characteristic scale needed to resolve this discreteness goes
as $1/\beta$. Hence, continuous non-locality emerges strictly in the limit of zero
temperature. We summarize the structure of the modular Hamiltonian in table~\ref{tab:summary}.

\begin{table}[h]
  \caption{Summary of our results for the modular Hamiltonian in different spin
    sectors. The definitions for $K_{\text{loc}}$ and $K^\pm_{\text{bi-loc}}$ are
    in~\eqref{eq:k-loc-23}--\eqref{eq:k-biloc}. The local and non-local terms at low
    temperature ($\beta\to\infty$) are given in~\eqref{eq:beta-large-beta}
    and~\eqref{eq:k-nonloc-large-beta}.}
  \label{tab:summary}
\begin{ruledtabular}
    \begin{tabular}{c c c c}
      $\nu$ &$\beta\to\infty$  &$\beta$ finite &$\beta\to 0$ \\\hline
      2     &local + cont. non-local &$K_{\text{loc}}+K^+_{\text{bi-loc}}$
                                               &$\beta\ii\partial_x\delta(x-y)$ \\
      3     &local &$K_{\text{loc}}+K^-_{\text{bi-loc}}$
                                               &$\beta\ii\partial_x\delta(x-y)$
    \end{tabular}
  \end{ruledtabular}
\end{table}

For multiple intervals, the only difference is that~\eqref{eq:mu-roots} now possesses one solution \emph{per interval} for a given
$k$, including the non-trivial ($x\neq y$) solutions for
$k=0$. In the low temperature limit, these extra terms yield precisely the
bi-local terms of~\cite{Casini:2009vk,Klich:2015ina}.

 During the final stage of this project, related results were independently
  reported in~\cite{Hollands:2019hje} and~\cite{Blanco:2019xwi}. Eqs.~(145) and~(146) of~\cite{Hollands:2019hje} give the modular flow of the
  correlator. The generator of this flow corresponds to the expectation
  value of our result for the modular Hamiltonian. Finally, the versatility of the resolvent method has allowed to compute the associated entanglement entropy\,\cite{Fries:2019acy}, and can also be used to study other quantities related to the entanglement spectrum.

\begin{acknowledgments}
\section{Acknowledgments}
We are very grateful to D. Blanco and G. P\'erez-Nadal for collaboration in the initial
stages of this project.

We thank J. Camps, B. Czech, M. Heller, H. Hinrichsen, C. Northe and G. Wong for helpful
discussions on the subject. PF is financially supported by the DFG project DFG HI
744/9-1. The Gravity, Quantum Fields and Information group at AEI is generously supported
by the Alexander von Humboldt Foundation and the Federal Ministry for Education and
Research through the Sofja Kovalevskaja Award. IR also acknowledges the hospitality of
Perimeter Institute, where part of this work was done.
\end{acknowledgments}

\section{Supplemental material}

\subsection{Conventions}

We work with the Dedekind eta and Jacobi theta functions defined as:
\begin{align*}
  \eta(q^2) &:= q^{1/12} \prod_{k\geq 1}(1-q^{2k}), \\
  \vartheta_3(w|q) &:= \sum_{k\in\zz} q^{k^2} w^{2k}, \\
  \vartheta_4(w|q) &:= \vartheta_3(\ii w|q), \\
  \vartheta_2(w|q) &:= q^{1/4} w \vartheta_3(\sqrt q w|q), \quad \text{and} \\
  \vartheta_1(w|q) &:= -\ii q^{1/4} w \vartheta_3(\ii \sqrt q w|q).
\end{align*}

  \subsection{The resolvent from the propagator}
  Let us give a quick derivation of \eqref{eq:mod-ham-int}. The basic idea is that given a
  holomorphic function $f(z)$ and an operator $\mathcal{G}$, we can determine
  $f(\mathcal{G})$ by using Cauchy's integral formula for each of the eigenvalues of
  $\mathcal{G}$.
  
  In our case, we wish to find $K=-\log [G^{-1}-1]$. Since the spectrum of
  $K$ is real, this equation tells us that the spectrum of the propagator $G$ is a subset
  of the open interval $(0,1)$. Consider now a specific eigenvalue $g$ of $G$. We can
  use Cauchy's integral formula to find
  \begin{align*}
    \log [g^{-1}-1]
    &=\log [1-g] - \log g\\
    &=\frac 1{2\pi \ii} \oint_{\gamma_g} \!\dif z \bigg[\frac 1{z-1+g} - \frac 1{z-g}\bigg] \log z,
  \end{align*}
  where $\gamma_g$ is a contour that encircles both $g$ and $1-g$. Notice that $\log z $
  possesses a branch cut, which we choose to place along the negative real axis -- the
  contour $\gamma_g$ must not intersect this cut. Then, since the function is holomorphic
  everywhere else in the plane, we can freely deform $\gamma_g$ such that integration
  only has to be done along the branch cut and a circle at infinity. The latter
  contribution however vanishes since the integrand is bounded by $z^{-2} \log z$ for
  large $z$, and we obtain
  \[
    \frac 1{2\pi\ii} \int_{-\infty}^0 \!\dif z \bigg[\frac 1{z-1+g} - \frac 1{z-g}\bigg] \\
    \times \Big[\log^+ z - \log^- z\Big],
  \]
  with $\log^\pm$ referring to the values just above and below the branch cut. After evaluating
  the last bracket to $2\pi\ii$, we can change variables to
  $\xi = 1/2 -z$ to find
  \[
    \log [g^{-1}-1]
    = \int_{1/2}^\infty \!\dif \xi \bigg[\frac 1{g-\xi-1/2}+\frac 1{g+\xi-1/2}\bigg].
  \]
  As this formula is valid for every eigenvalue $g$ of $G$, it also holds as an operator
  statement, hence we arrive at~\eqref{eq:mod-ham-int}.

\subsection{Deriving the resolvent}
In this section, we derive the solution $F^\nu_V$ (given in \eqref{eq:solution}) to the singular
integral equation~\eqref{eq:singular-int-eq-rad}. Let us start with the
functions~\cite{Klich:2015ina}
\begin{equation}
  \label{eq:circle-omega}
  \omega_n(w) :=
  \frac{\sin (\pi(a_n-z))}{\sin (\pi(b_n-z))} =
  \frac{\beta_n}{\alpha_n} \frac{\alpha_n^2-w^2}{\beta_n^2-w^2},
\end{equation}
which provide the correct branch-cut on the cylinder. Choosing the branch cut of the
logarithm along the negative real line, we see that
\begin{equation}
  \label{eq:branch-cut-circle}
  \prod_{n=1}^N \frac{\omega_n^{\ii h}(w \ee^{-\epsilon''})}
  {\omega_n^{\ii h}(w \ee^{\epsilon''})} = \ee^{2\pi h}
\end{equation}
for $w \in A$. Note that~\eqref{eq:circle-omega} is not well defined on the torus since it
transforms non-trivially under $w \to qw$. We shall remedy this by defining
\begin{equation}\label{eq:big-omega}
  \log \Omega_n(w) :=
  \sum_{k\in\zz} \big[\log(\omega_n(q^k w)) - \log(\omega_n(q^k))\big],
\end{equation}
where the second term in the brackets is to ensure absolute convergence. We made the
logarithms explicit in order not to break the behaviour~\eqref{eq:branch-cut-circle} at
the branch cut.

At first sight, $\Omega(w)$ seems doubly-periodic: by construction, $\omega$ is periodic
with respect to the spatial circle, and now we sum over all translations along imaginary
time. However, $\Omega(w)$ has a non-vanishing residue within each fundamental region due
to the branch-cut, and thus cannot be elliptic. Instead, it turns out to be
quasi-periodic, as is seen by putting a cutoff in the sum~\eqref{eq:big-omega}, and then
computing $\Omega(q w)$. Then, the series acquires a prefactor originating in
\begin{equation*}
  \lim_{k \to \pm \infty} \omega_n(q^k w) = \bigg[\frac{\beta_n}{\alpha_n}\bigg]^{\mp 1}.
\end{equation*}
This yields the quasi-periodicity
\begin{equation*}
  \prod_{n=1}^N \Omega_n^{\ii h}(q w) = \lambda^2 \prod_{n=1}^N \Omega_n^{\ii h}(w),
\end{equation*}
where $\lambda$ is defined in~\eqref{eq:def-lambda}. To cancel off the quasi-periodicity
and to introduce the desired pole, we multiply with a combination of theta functions. We
find
\begin{align}
  \label{eq:solution-cpl-plane}
  F^\nu_V(\xi;u,v)
  &= \frac{\eta^3(q^2)}{\ii \vartheta_1(uv^{-1}\ee^{\epsilon'}|q)}
    \frac{\vartheta_\nu(\lambda uv^{-1}|q)}{\vartheta_\nu(\lambda|q)} \nonumber \\
  &\hspace{2.5cm} \times \prod_{n=1}^N\frac{\Omega_n^{\ii h}(u \ee^{\epsilon''})}
    {\Omega_n^{\ii h}(v\ee^{-\epsilon''})},
\end{align}
where we made our choice of branches in $\Omega^{\ii h}_n$ explicit (our choice is such
that the residue evaluation in the main body of the paper does not cross the branch
cut). Eq.~\eqref{eq:solution-cpl-plane} now solves~\eqref{eq:singular-int-eq-rad}.

Finally, let us rewrite this in terms of more familiar elliptic functions. Note
that~\eqref{eq:branch-cut-circle} implies
\begin{equation*}
  \prod_{n=1}^N\frac{\Omega_n^{\ii h}(u \ee^{\epsilon''})}
  {\Omega_n^{\ii h}(v\ee^{-\epsilon''})}
  = \ee^{-2\pi h} \prod_{n=1}^N\frac{\Omega_n^{\ii h}(u \ee^{\epsilon''})}
  {\Omega_n^{\ii h}(v\ee^{\epsilon''})}
\end{equation*}
and, now that the numerator and denominator are on the same side of the branch cut, we can move
the product into the complex power to find
\begin{equation*}
  \prod_{n=1}^N\frac{\Omega_n^{\ii h}(u \ee^{\epsilon''})}{\Omega_n^{\ii h}(v\ee^{-\epsilon''})}
  = \ee^{-2\pi h}
  \bigg[\prod_{n=1}^N \prod_{k \in \zz} \frac{\omega_n(q^k u \ee^{\epsilon''})}
  {\omega_n(q^k v \ee^{\epsilon''})}\bigg]^{\ii h}
\end{equation*}
for $u,v \in A$.  After some algebra and an application of the Jacobi triple
product~\cite{Whittaker:2009zvh}, this simplifies the
solution~\eqref{eq:solution-cpl-plane} to~\eqref{eq:solution} with $\Omega_V$
from~\eqref{eq:def-omega}.

\subsection{Deriving the modular Hamiltonian}
In this section, we provide the mains steps to evaluate the integral
expression~\eqref{eq:mod-ham-int-2} for the modular Hamiltonians. We restrict to purely
imaginary $\tau = \ii \beta$---the general case can be restored by analytic
continuation. Let us first change the variable of integration from $\xi$ to
\begin{equation*}
  \Lambda := \lambda^2 = \ee^{2\pi L h} = \bigg[\frac{\xi+1/2}{\xi-1/2}\bigg]^L,
\end{equation*}
such that~\eqref{eq:mod-ham-int-2} turns into
\begin{equation*}
  K^\nu_V = \frac 1L \int_0^\infty \frac{\dif \Lambda}\Lambda
  \frac{\eta^3(q^2)}{\ii \vartheta_1(uv^{-1}|q)}
  \frac{\vartheta_\nu(\sqrt\Lambda uv^{-1}|q)}{\vartheta_\nu(\sqrt\Lambda|q)} \Lambda^{i \mu},
\end{equation*}
where we use the shorthand notation
\begin{equation*}
  \mu := \frac 1{2\pi L} \log \frac{\Omega_V(u)}{\Omega_V(v)}.
\end{equation*}

To evaluate the above integral, note the following two facts:
  \begin{itemize}
  \item Since we merged the two occurences of $F^\nu_V(\pm\xi)$ in~\eqref{eq:mod-ham-int-2}
    into a single integral, integration has to be done symmetrically with respect to
    $\xi \leftrightarrow -\xi$, i.e., $\Lambda \leftrightarrow \Lambda^{-1}$.
  \item The integrand is oscillatory for $\Lambda \to 0, \infty$.
  \end{itemize}
  
This requires that we introduce a symmetric regulator
$r_\epsilon(\Lambda) = r_\epsilon(\Lambda^{-1})$ to tame the integral, allowing us to
evaluate it via standard complex analysis methods. We choose
\begin{equation}
  \label{eq:def-reg}
  r_\epsilon(\Lambda) := \frac{(1+\epsilon)^2}{(\Lambda+\epsilon)(\Lambda^{-1}+\epsilon)}
\end{equation}
to otain
\begin{align}
  \label{eq:mod-ham-int-3}
  K^\nu_V
  &= \lim_{\epsilon \searrow 0} \frac 1L \int_0^\infty \dif \Lambda
    \frac{(1+\epsilon)(1+\epsilon^{-1})}{(\Lambda+\epsilon)(\Lambda+\epsilon^{-1})}
    \nonumber \\
  &\hspace{1.5cm}\times \frac{\eta^3(q^2)}{\ii \vartheta_1(uv^{-1}|q)}
    \frac{\vartheta_\nu(\sqrt\Lambda uv^{-1}|q)}{\vartheta_\nu(\sqrt\Lambda|q)} \Lambda^{\ii \mu}.
\end{align}

\begin{figure}[h]
\begin{center}
\includegraphics[width=7.1cm]{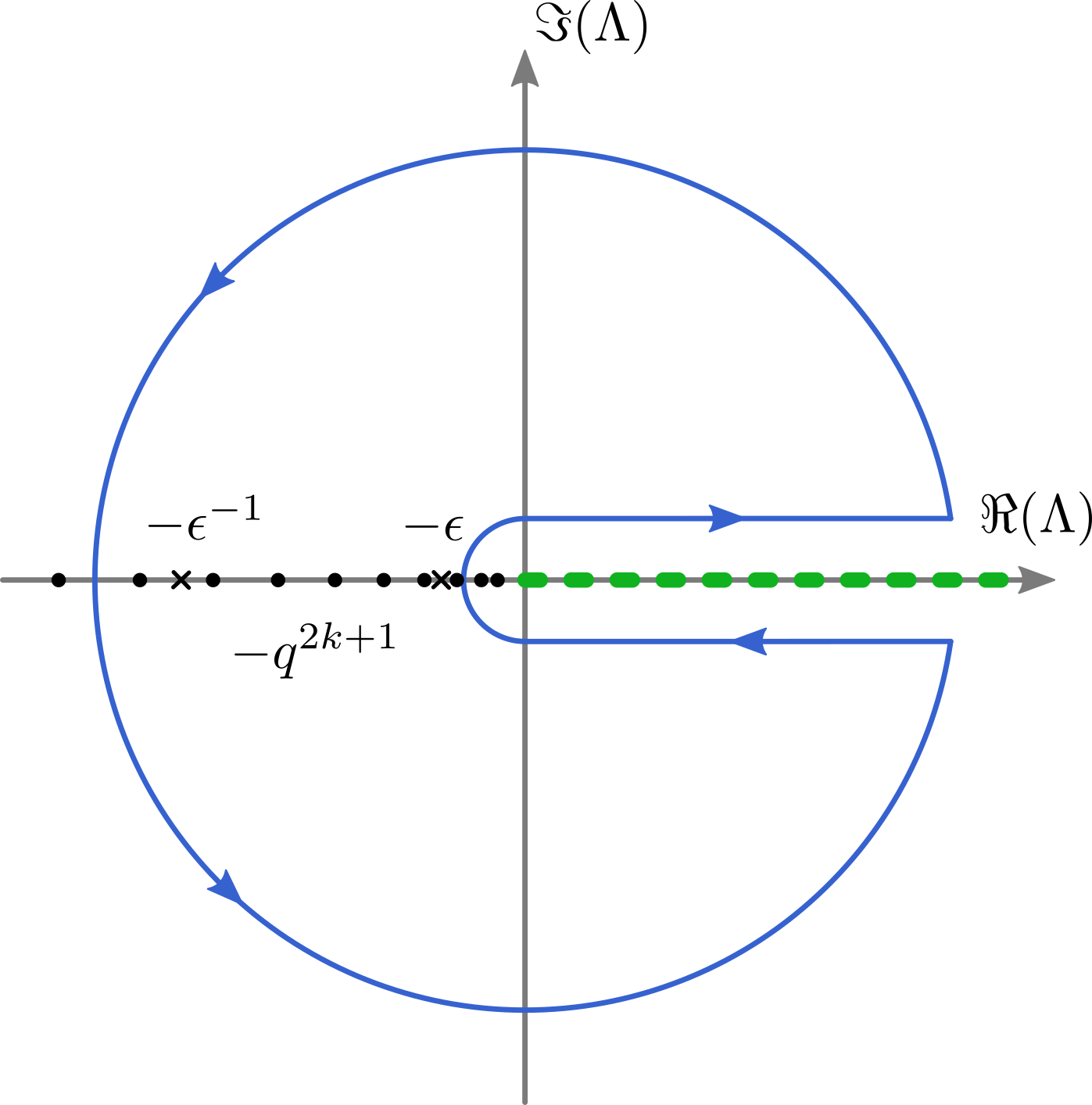}
\caption{Contour for the integral~\eqref{eq:pacman}. The integral along
  the blue solid line is equal to the sum of all residues at $\Lambda \to -q^{2k+1}$
  (black dots) and at $\Lambda \to -\epsilon^{\pm 1}$ (black crosses). The contour avoids
  the branch cut along the positive real axis (green dashed line).}
\label{fig:pacman} 
\end{center}
\end{figure}

The integral~\eqref{eq:mod-ham-int-3} can now be evaluated using contour integration. To this
end, consider the integral
\begin{align}
  \label{eq:pacman}
  I^\nu_\epsilon
  &:= \frac 1L \oint_\gamma \dif \Lambda
    \frac{(1+\epsilon)(1+\epsilon^{-1})}{(\Lambda+\epsilon)(\Lambda+\epsilon^{-1})} \nonumber\\
  &\hspace{2cm}\times \frac{\eta^3(q^2)}{\ii \vartheta_1(uv^{-1}|q)}
    \frac{\vartheta_\nu(\sqrt\Lambda uv^{-1}|q)}{\vartheta_\nu(\sqrt\Lambda|q)} \Lambda^{\ii \mu},
\end{align}
where the contour $\gamma$ is as depicted in Fig.~\ref{fig:pacman}.

The circular contributions vanish due to the falloff of the regulator. Choosing the branch
cut of $\Lambda^{\ii \mu}$ along the positive real axis, we see that two remaining
horizontal contributions yields two almost identical terms, differing differ only by a
global prefactor of $-\ee^{-2\pi \mu}$. We thus find
\begin{equation}
  \label{eq:mod-ham-i}
  \lim_{\epsilon \searrow 0} I^\nu_\epsilon = (1-\ee^{-2\pi \mu}) K^\nu_V.
\end{equation}
By Cauchy's theorem, $I^\nu_\epsilon$ can also be expressed as a sum over residues,
yielding a series expression for $K^\nu_V$. We shall do this explicitly for $\nu = 3$ and
briefly mention the differences for $\nu = 2,4$ at the end.

The poles of the integrand are of two types (see Fig.~\ref{fig:pacman}): Two come from the
regulator~\eqref{eq:def-reg}, located at $\Lambda\to-\epsilon$ and
$\Lambda\to-\epsilon^{-1}$. The other (infinitely many) poles come from the poles of the
`propagator-like' term. As can be seen from either the Laurent expansion of this term (see
section below) or directly from the Jacobi triple product, we have the leading divergences
\begin{equation*}
  \frac{\eta^3(q^2)}{\ii \vartheta_1(uv^{-1}|q)}
  \frac{\vartheta_3(\sqrt\Lambda uv^{-1}|q)}{\vartheta_3(\sqrt\Lambda|q)} \sim
  -\frac{(q^{-1}uv^{-1})^{-2k-1}}{\Lambda + q^{2k+1}}
\end{equation*}
at $\Lambda \to -q^{2k+1}$ for $k\in \zz$. Keeping in mind that the negative sign of poles
always has to be written as $\ee^{+\ii \pi}$ due to our choice of branch cut, this yields
\begin{align}
  K^3_V
  &= \frac{2\pi\ii}L \frac 1{\ee^{\pi \mu}-\ee^{-\pi \mu}} \lim_{\epsilon \searrow 0}
    \bigg[\frac{\eta^3(q^2)}{\ii \vartheta_1(uv^{-1}|q)} \nonumber\\
  &\hspace{.8cm}\times \bigg(
    \frac{\vartheta_4(\sqrt\epsilon uv^{-1}|q)}{\vartheta_4(\sqrt\epsilon|q)}
    \epsilon^{\ii\mu} - (\epsilon \to \epsilon^{-1})\bigg) \nonumber\\
  &\hspace{2cm}+ \sum_{k\in\zz}
    \frac{(uv^{-1}q^{-\ii\mu})^{-2k-1}}{(q^{2k+1}-\epsilon)(q^{-2k-1}-\epsilon)} \bigg].
    \label{eq:mod-ham-series}
\end{align}

Let us have a look at the series in the last line: Using the Laurent expansions below,
this can be rewritten as
\begin{equation*}
  \frac{\eta^3(q^2)}{\ii \vartheta_1(uv^{-1} q^{-i\mu} |q)}
  \frac{\vartheta_4(\sqrt\epsilon uv^{-1} q^{-i\mu} |q)}{\vartheta_4(\sqrt\epsilon|q)}
  - (\epsilon\rightarrow \epsilon^{-1}).
\end{equation*}

We choose the cutoff to be $\epsilon = q^{2m}$ with very large $m\in\mathbb Z$ to avoid
the poles at $q^{2k+1}$, so that we only deal with simple poles. Then, putting everything
together into~\eqref{eq:mod-ham-series} and using the quasiperiodicities of $\vartheta_4$,
one finds
\begin{equation*}
  K^3_V(x,y)
  = \lim_{m \to \infty} P(x,y) \sin \big(2m\pi(x-y+\beta \mu)\big)
\end{equation*}
with
\begin{align*}\label{}
  P(x,y)
  &= \frac{2\pi}{L\sinh \pi\mu(x,y)} \bigg[
    \frac{\eta^3(q^2)}{\ii \vartheta_1(uv^{-1}|q)}
    \frac{\vartheta_4(uv^{-1}|q)}{\vartheta_4(1|q)} \nonumber\\
  &\hspace{1.5cm}- \frac{\eta^3(q^2)}{\ii \vartheta_1(uv^{-1}q^{-i\mu}|q)}
    \frac{\vartheta_4( uv^{-1}q^{-i\mu}|q)}{\vartheta_4(1|q)} \bigg].
\end{align*}
As already stated above, this limit must be understood in the sense of distributions.

We see that $K^3_V$ contains essentially two factors: the term involving the sine function
is highly oscillatory for $m\to\infty$, except at solutions of
\begin{equation}
  \label{eq:mu-roots-2}
  x-y+\beta\mu(x,y) = k\in\mathbb Z.
\end{equation}
As a distribution, it vanishes when integrated against any regular test function. However,
the remaining factor $P(x,y)$ is not regular since it has poles, and thus we must examine
its behaviour in their vicinity, which will lead to finite contributions. These poles
coincide precisely with the solutions to~\eqref{eq:mu-roots-2}, which are of two kinds:
the trivial solution $x=y$ will lead to a local term, while the other solutions $x\neq y$
will give bi-local contributions. Let us start with the latter.

Close to these solutions, a straightforward calculation shows that
\begin{equation*}
  P(x,y) \sim \frac{\ii\pi}{L\sinh \pi\mu(x,y)} \frac 1{\sin (\pi [x-y+\beta\mu(x,y)])}.
\end{equation*}
Combined with the oscillatory term, we recognize the Dirichlet kernel~\cite{Rudin:1976ksh}
representation of the anti-periodic Dirac delta
\begin{equation}
  \label{eq:dirichlet}
  \lim_{m\to\infty} \frac{\sin 2m\pi z}{\sin\pi z} = \sum_{k\in \zz} (-1)^k \delta(z-k),
\end{equation}
yielding the final expression for the modular Hamiltonian for $x\neq y$,
\begin{equation}
  \label{eq:biloc}
  \frac{\ii\pi}{L\sinh\pi\mu} \sum_{k\in \zz} (-1)^k \delta(x-y+\beta\mu(x,y)-k).
\end{equation}

Now we turn to the solution $x=y$, which is special as it leads to a second order pole in
$P$. Similarly to before, in the vicinity of that solution, $P(x,y)$ takes the form
\begin{equation*}
  -\frac{\ii\beta}L \frac 1{x-y}\frac 1{\sin (\pi[x-y+\beta\mu(x,y)])}, 
\end{equation*}
which together with the oscillatory term leads to 
\begin{equation}
  \label{eq:sing-frac}
  -\frac{\ii\beta}L \frac{\delta(x-y+\beta \mu(x,y))}{x-y}.
\end{equation}

Note that we did not need to consider the terms with $k\neq 0$ as in~\eqref{eq:dirichlet}
since we only deal with the solution $x=y$. As a last step, we use the methods
from~\cite{Casini:2009vk} to rewrite the singular fraction~\eqref{eq:sing-frac} as
\begin{equation}
  \label{eq:loc}
  \frac{\beta}{L} \frac{[\ii\partial_x + f(x)] \delta(x-y)}{1+\beta (\partial_x\mu)(y,y)},
\end{equation}
where $f(x)$ is fixed by hermiticity.

Now we focus on the case of a single interval. Again we begin by considering on the
bi-local terms. Since $\mu(x,y)$ is monotonically increasing with respect to $x$ in the
interval, eq.~\eqref{eq:mu-roots-2} has a unique solution for each $k\in\zz$. In
particular, note that for $k=0$, the solution is $x=y$. Since we already consider this
contribution separately in~\eqref{eq:loc}, we can explicitly exclude it from the
series~\eqref{eq:biloc}. The final expression for the modular Hamiltonian for a single
interval is then given by the sum of~\eqref{eq:k-loc-23} and~\eqref{eq:k-biloc}. Finally, replacing~\eqref{eq:k-biloc} in~\eqref{rho}, the bi-local modular Hamiltonian takes the form \eqref{eq:Kpmbiloc}.

An analogous calculation holds for $\nu=2$, with one small adjustment: Since
  the poles of the Laurent expansion are instead located at $-q^{2k}$, we obtain the
  periodic version of the Dirichlet kernel in~\eqref{eq:dirichlet}. The rest of the
  calculation is identical.

\section{Laurent expansion}
To better understand the location and behaviour of the poles of the ``propagator-like''
terms in~\eqref{eq:solution}, we derived their Laurent expansions. The coefficients may be
computed as contour integrals which vastly simplify due to the quasi-periodicities of the
theta functions. In the fundamental domain $|q|^{1/2} < |w| < |q|^{-1/2}$, the result then
takes the form of Lambert series
\begin{align*}
  &\frac{\eta^3(q^2)}{\ii \vartheta_1(w|q)}\frac{\vartheta_3(\lambda w|q)}
    {\vartheta_3(\lambda|q)} = \frac 1{w-w^{-1}} \\
  &\hspace{3cm}+ \sum_{\substack{k\geq 1\\k\text{ odd}}} \left[\frac{w^kq^k}
  {\lambda^{-2}+q^k}-\frac{w^{-k}q^k}{\lambda^2+q^k}\right], \\
  &\frac{\eta^3(q^2)}{\ii \vartheta_1(w|q)}\frac{\vartheta_4(\lambda w|q)}
    {\vartheta_4(\lambda|q)} = \frac 1{w-w^{-1}} \\
  &\hspace{3cm}- \sum_{\substack{k\geq 1\\k\text{ odd}}} \left[\frac{w^kq^k}
  {\lambda^{-2}-q^k}-\frac{w^{-k}q^k}{\lambda^2-q^k}\right], \\
  &\frac{\eta^3(q^2)}{\ii \vartheta_1(w|q)}\frac{\vartheta_2(\lambda w|q)}
    {\vartheta_2(\lambda|q)} =
    \frac 12\frac{w+w^{-1}}{w-w^{-1}} + \frac 12 \frac{\lambda^2-1}{\lambda^2+1} \\
  &\hspace{3cm}+ \sum_{\substack{k\geq 2\\k\text{ even}}} \left[\frac{w^kq^k}
  {\lambda^{-2}+q^k}-\frac{w^{-k}q^k}{\lambda^2+q^k}\right].
\end{align*}

\bibliography{Refs}
\bibliographystyle{ieeetr}

\end{document}